\input{aipcheck}

\documentclass[
    ,final            
  ]
  {aipproc}

\layoutstyle{6x9}

\begin{document}

\title{Cosmological singularities and modified theories of gravity}


\author{Leonardo Fern\'andez-Jambrina}{
  address={Matem\'atica Aplicada, E.T.S.I. Navales, Universidad
Polit\'ecnica de Madrid,\\
Arco de la Victoria s/n,  E-28040 Madrid, Spain}
}

\author{Ruth Lazkoz}{
  address={F\'\i
sica Te\'orica, Facultad de Ciencia y Tecnolog\'\i a, Universidad del
Pa\'\i s Vasco, \\ Apdo.  644, E-48080 Bilbao, Spain}
}

\begin{abstract}
We consider perturbative modifications of the Friedmann equations in
terms of energy density corresponding to modified theories of gravity
proposed as an alternative route to comply with the observed
accelerated expansion of the universe.  Assuming that the present
matter content of the universe is a pressureless fluid, the possible
singularities that may arise as the final state of the universe are
surveyed.  It is shown that, at most, two coefficients of the
perturbative expansion of the Friedman equations are relevant for the
analysis.  Some examples of application of the perturbative scheme are
included.
\end{abstract}

\maketitle


\section{Introduction}

Recent astronomical data from Type Ia supernovae \cite{Davis:2007na}
as well as from the CMB spectrum \cite{wmap} confirm that our universe
is undergoing an accelerated expansion period.  In order to comply
with this feature, one may resort to postulating a dark energy content
for the universe \cite{Padmanabhan:2006ag}, with undesired properties,
such as violation of some energy conditions, or going beyond general
relativity in the quest for another theory of gravity
\cite{Maartens:2007,Durrer:2007re,salvc,disc}.

With either approaches for dealing with the observed accelerated
expansion, cosmology is much richer than it was thought in the
previous century. According to classical cosmologies, the universe 
started at an initial singularity, the Big Bang, and it was doomed to 
expand forever, since the matter content was not dense enough to stop 
expansion and collapse into a final singularity.

However, accelerated expansion of the universe has lead to consider 
other plans for the end of the universe in the form of Big Rip 
singularities \cite{Caldwell:2003vq}, directional singularities 
\cite{direct} or milder sudden singularities 
\cite{Barrow:2004xh}. It is therefore interesting to tell under which 
circumstances a modified theory of gravity may lead to such fate for 
the universe.

\section{Modified gravity}

Instead of dealing with a full theory of modified gravity, we focus on
the consequences at the level of cosmological equations, namely
Friedmann equation, just requiring that it admits a generalised power
expansion on the density $\rho$ around a value $\rho_{*}$,
\begin{equation}
\left(\frac{\dot a}{a}\right)^{2}=H^{2}=h_{0}(\rho-\rho_{*})^{\xi_{0}}
+h_{1}(\rho-\rho_{*})^{\xi_{1}}+\cdots,\quad \xi_{0}<\xi_{1}<\cdots.\label{pert}
\end{equation}

The standard density term arises as the lineal term with an 
exponent equal to one and the cosmological constant appears with a null 
exponent in this expansion. Further terms are interpreted as 
modifications of the theory.

On the other hand, the energy conservation law implies
\begin{equation}
\dot \rho+ 3H(\rho+p)=0,\label{cons}
\end{equation}
but assuming that the accelerated expansion  of the universe is 
solely due to the modification of the theory of gravity, we choose a 
pressureless dust as matter content of the universe, so that  the 
scale factor of the universe and the density are related by $\rho 
a^3=K$.

Thereby we may get rid of the scale factor and write down a modified Friedmann 
equation in terms of density:
\begin{eqnarray}
\frac{\dot\rho}{\rho}&=&
-3\sqrt{h_{0}}(\rho-\rho_{*})^{\xi_{0}/2}-
\frac{3}{2}\frac{h_{1}}{\sqrt{h_{0}}}(\rho-\rho_{*})^{\xi_{1}-\xi_{0}/2}
+\cdots\quad\,\label{conspert}.
\end{eqnarray}

Solving this equation provides a perturbative expansion of the
density in coordinate time, which we want to compare with a similar expansion for the scale factor:
\[    
a(t)=c_{0}|t-t_{0}|^{\eta_{0}}+c_{1}|t-t_{1}|^{\eta_{1}}+\cdots,\quad
\eta_{0}<\eta_{1}<\cdots, \]in terms of coordinate time.  This is
useful, since in \cite{puiseux} we have related the exponents
$\eta_{i}$ with the strength of singularities according to the
standard definitions by Tipler \cite{tipler} and Kr\'olak
\cite{krolak}, as we see in Table 1.  The strength of the
singularities just points out if tidal forces are strong enough to
disrupt finite objects on crossing them.

\begin{table}[h]
   \begin{tabular}{cccccc}
   \hline
   ${\eta_{0}}$ & ${\eta_{1}}$ & $\eta_{2}$ &\textbf{Tipler} &
   \textbf{Kr\'olak} & \textbf{N.O.T.} \\
   \hline
   $(-\infty,0)$ & $(\eta_{0},\infty)$ &   $(\eta_{1},\infty)$ &
   Strong & Strong  & I\\ 
   $0$ & $(0,1)$ &   $(\eta_{1},\infty)$ &   Weak & Strong & III \\
      & $1$ & $(1,2)$ & Weak & Weak & II \\
      &  & $[2,\infty)$ & Weak & Weak & IV \\
      & $(1,2)$ &  $(\eta_{1},\infty)$ & Weak & Weak & II \\
      & $[2,\infty)$ &   $(\eta_{1},\infty)$ & Weak & Weak &  IV \\
   $(0,\infty)$ & $(\eta_{0},\infty)$ &   $(\eta_{1},\infty)$ &
Strong & Strong & Crunch\\
   \hline
   \end{tabular}
\caption{Singularities in  cosmological models}
\end{table}

The last column refers to the classification of future singularities
in \cite{Nojiri:2005sx}):

\begin{itemize}
   \item  Type I: ``Big Rip'': divergent $a$.

   \item  Type II: ``Sudden'': finite $a$, $H$, divergent $\dot H$.

   \item  Type III: ``Big Freeze'': finite $a$, divergent $H$.

   \item  Type IV: ``Big Brake'': finite $a$, $H$, $\dot H$, but
   divergent higher derivatives.
\end{itemize}

We have chosen the decreasing density branch of the Friedmann
equation, since we wish to mimic the actual expansion phase of the
universe.  There are two possibilities: either there is a critical 
value $\rho_{*}$ for which the density stops decreasing or it goes 
on decreasing without a lower bound. The results are consigned in 
tables 2 and 3. More details may be found in \cite{fate}.

Singularities in models with null critical density are determined by 
the sign of the first exponent $\xi_{0}$. If it is positive, 
expansion goes on forever and no singularity arises, but if the 
exponent is negative, a Big Rip singularity comes up due to the 
accelerated expansion.

\begin{table}[h]
   \begin{tabular}{cccc}
   \hline
   ${\xi_{0}}$ &  \textbf{Tipler} &
   \textbf{Kr\'olak} & \textbf{N.O.T.} \\\hline
   $(-\infty,0)$ &
   Strong & Strong  & I\\ 
   $[0,\infty)$ &
Non-singular & Non-singular & Non-singular\\
   \hline
   \end{tabular}
\caption{Singularities in models with $\rho_{*}=0$}

\end{table}

On the contrary, the structure of models with a finite critical 
density $\rho_{*}$ is much richer. No Big Rip singularities appear 
and they are all weak. Therefore, they cannot be considered the final 
stage of the universe \cite{suddenferlaz}.

Models with  negative $\xi_{0}$ exponent 
 have divergent $H$, which is a Big Freeze singularity, 
which is weak under Tipler's definition, but strong with Kr\'olak's.

Models with null $\xi_{0}$ exhibit a cosmological constant term and 
produce just weak singularities. The second exponent $\xi_{1}$ can be 
used to tell the derivative of $H$ which is singular, but none of 
these can be taken as the final fate of the universe, due to the 
weakness of the singularity.

Finally, models with no cosmological constant, $\xi_{0}>0$ show the 
same types of weak singularities.

Examples of proposed models belonging to these families may be found 
in \cite{fate}.

\begin{table}[h]
   \begin{tabular}{ccccc}
   \hline
   ${\xi_{0}}$ & ${\xi_{1}}$ & \textbf{Tipler} &
   \textbf{Kr\'olak} & \textbf{N.O.T.} \\
   \hline
   $(-\infty,0)$ & $(\xi_{0},\infty)$ &
   Weak & Strong  & III\\ 
   $0$ & $(0,1)$ &      Weak & Weak & II \\
    & $[1,\infty)$ &  Weak & Weak & IV \\
   $(0,1)$ & $(\xi_{0},\infty)$ &
Weak & Weak & II\\ 
   $[1,2)$ & $(\xi_{0},\infty)$ &
Weak & Weak & IV\\
   \hline
   \end{tabular}
\caption{Singularities in models with $\rho_{*}\neq0$}
\end{table}

\begin{theacknowledgments}
L.F.-J. is supported by the Spanish Ministry
of Education and Science Project FIS-2005-05198.  R.L. is supported by
the University of the Basque Country through research grant
GIU06/37 and by the Spanish Ministry of Education and
Culture through research grant FIS2007-61800. \end{theacknowledgments}



\end{document}